\documentstyle[12pt]{article}
\def\Eq{\begin{equation}}	\def\End{\end{equation}}
\def\Eqa{\begin{eqnarray}}	\def\Enda{\end{eqnarray}}
\def\Endl#1{\label{#1} \End}	\def\Endla#1{\label{#1} \Enda}
\def\puteq#1{eq.~(\ref{#1})}	\def\Puteq#1{Eq.~(\ref{#1})}
\def\ie{{\it i.e.}}		\def\ord#1{{\cal O}({#1})}
\def\gst{g_{\Sigma\Lambda}}	\def\eps{\epsilon}
\def\eg{{\it e.g.}}
\def\slash#1{\,\raise.15ex\hbox{$/$}\mkern-10.mu #1}
\def\et#1#2{\eta_{\ifnum#1>0 #1 \fi}^{(#2)}}
\def\frac#1#2{{\textstyle {#1 \over #2}}}

\input epsf
\def\boxit#1{\vbox{\hrule\hbox{\vrule{#1}\vrule}\hrule}}
\def\INSERTFIG#1#2#3{\epsfxsize=#1in \begin{center}
  \hbox to\hsize{\hfil\boxit{\epsffile{#2}}\hfil} {\sl #3} \end{center}}

\begin{document}
\begin{titlepage}
\begin{center}
March, 1996	\hfill CMU-HEP95-21\\
{}~		\hfill DOE/ER/40682-109\\
{}~		\hfill hep-ph/9511309\\
\vskip1.0in {\Large\bf
% Journal layout - remove preprint numbers above and use:
% ~\vskip.25in {\Large\bf
Large-$N_c$ Relations Among Isgur-Wise Functions}
\vskip.3in
David E. Brahm\footnote{Email: \tt brahm@fermi.phys.cmu.edu} and
James Walden\footnote{Email: \tt walden@fermi.phys.cmu.edu}\\~\\
{\it Carnegie Mellon Physics Dept., Pittsburgh PA 15213}
\end{center}

\vskip.5in
\begin{abstract}
We investigate the relations that must hold among baryonic Isgur-Wise functions
$\eta_i$ in the large-$N_c$ limit from unitarity constraints, and compare to
those found by Chow using the Skyrme model [or $SU(4)$].  Given the exponential
dropoff of the $\eta_i$ away from threshold, unitarity requires only that the
usual normalization conditions hold at $w=1$, and that $\eta=\eta_1$ near
threshold.  Our results are consistent with, but less powerful than, the Skyrme
model relations.
\end{abstract}
\end{titlepage}
% Journal layout:
% \Lspace{1.8}

% ======================= TEXT BEGINS ==========================

\section{Introduction}
QCD simplifies greatly in the chiral \cite{chiral}, heavy-quark \cite{heavy},
and large-$N_c$ \cite{bign} limits (where $N_c$ is the number of colors).  The
Skyrme \cite{sky1} model of heavy baryons \cite{wit,skyrme,jmw} incorporates
all of these, and makes powerful predictions \cite{chow} about the baryonic
Isgur-Wise functions $\eta$, $\eta_1$, and $\eta_2$.  Since it has been
conjectured that all parameter-independent Skyrme model predictions can be
derived just from large-$N_c$ unitarity constraints, we investigate what these
constraints alone can tell us about Isgur-Wise functions.

Our model-independent large-$N_c$ results turn out to be less powerful than
Chow's Skyrme model predictions.  Using the exponential form $\eta \sim
\exp\{-\lambda N_c^{3/2} (w-1)\}$ given by Jenkins, Manohar and Wise
\cite{jmw}, unitarity requires only that the usual normalization conditions
hold at $w=1$, and that $\eta=\eta_1$ near threshold.

\subsection{Baryon Isgur-Wise Functions}

The weak transition $\Lambda_b \to \Lambda_c\, W$ is characterized by a single
Isgur-Wise function, which represents the overlap of the spin-0 light degrees
of freedom (``brown muck''):
\Eq \langle L(v') | L(v) \rangle = \eta(w) \End
where $w\equiv v\cdot v'$.  $\Lambda_Q$ is an isospin singlet ($I \!=\! J \!=\!
0$).

The weak transition $\Sigma^{(*)}_b \to \Sigma^{(*)}_c\, W$ is characterized by
two other functions, since in this system the brown muck has spin 1:
\Eq \langle L_\nu(v') | L_\mu(v) \rangle = -\eta_1(w) g_{\mu\nu} + \eta_2(w)
  v'_\mu v_\nu \End
The $\Sigma_Q$ and $\Sigma^*_Q$ can be treated together in the heavy quark
limit, as a single ``superfield'' $\Sigma$ \cite{cho}, since they differ only
in the relative spin orientation of the heavy quark and the brown muck.
$\Sigma$ is an isospin triplet ($I \!=\! J \!=\! 1$).

The normalization of these functions at $w=1$ (``threshold'') is:
\Eq \eta(1)=\eta_1(1)=1 \Endl{norm}
Heavy quark symmetry makes no prediction for the value of $\eta_2(1)$.

Chow \cite{chow} found the following relations among baryon Isgur-Wise
functions using the Skyrme model:\footnote{Chow writes $(\zeta_1,\zeta_2)$ for
$(\eta_1,\eta_2)$ and uses ``east coast'' metric $g_{\mu\nu} = {\rm
diag}(-1,1,1,1)$.  Chow has recently derived the same relations from $SU(4)$
symmetry. \cite{chow2}}
\Eq \eta_1(w) = -(1+w)\eta_2(w) = \eta(w) \Endl{chowrel}
These relations are consistent with the normalizations in \puteq{norm},
and additionally predict that $\eta_2(1)=-1/2$.

\section{Diagrams}

\subsection{1-Loop Renormalization of $\eta(w)$}

In Fig.~1 we show the 1-loop renormalization (vertex and wavefunction) of
$\Lambda_b(v) \to \Lambda_c(v') W$ (\ie\ of $\eta$), which is calculated by Cho
\cite[eq.~(3.4)]{cho}. Since $(\gst/f)^2 \sim N_c$, the term that it multiplies
must vanish at least as fast as $1/N_c$.  The relevant piece is
\Eq \left[ 3\eta - (2r+w)\eta_1 + (w^2-1)\eta_2 \right] = \ord{1\over N_c},
  \qquad r \equiv {\ln\left(w + \sqrt{w^2-1}\right) \over \sqrt{w^2-1}}
  \Endl{1lp}

\INSERTFIG{6.49}{bign_1.eps}{Fig.~1: 1-loop renormalization of $\Lambda_b(v)
  \to \Lambda_c(v') W$}

One might be tempted to use the renormalization of $\Sigma\to \Sigma' W$ (\ie\
of $\eta_1$ and $\eta_2$), also calculated by Cho, to derive more relations.
However, in the large-$N_c$ limit there exists an $I \!=\! J$ tower of states
above the $\Lambda$ and $\Sigma$.  In particular, the state with $I \!=\! J
\!=\! 2$ contributes to the 1-loop renormalization of $\Sigma\to \Sigma' W$.
It introduces 3 new Isgur-Wise functions \cite[eq.~(2.26)]{falk}, only one of
which is normalized at $w=1$.  Thus no useful new information is obtained.

\subsection{Single Pion Emission}

In Fig.~2, we look at weak decay accompanied by single pion emission:
$\Lambda_b(v) \to \Sigma_c(v') \pi_l(q) W$.  The sum of the two diagrams gives
an invariant amplitude
\Eq {\cal M} = {\gst\over 2f}{g_2 V_{cb}\over 2\sqrt2} \left[ \bar
  \Sigma_{ij}^\mu \slash\epsilon^* (1-\gamma_5) \Lambda \right] \left[
  \epsilon^{ki} (T_l)^j_k + \epsilon^{kj} (T_l)^i_k \right] (F q_\mu + G
  v_\mu) \Endl{calM}
where
\Eq F \equiv {\eta\over v'\cdot q} - {\eta_1\over v\cdot q}, \qquad
    G \equiv \eta_2 {v'\cdot q \over v\cdot q} + \eta_1 - w \eta_2 \End
and the $T_l$'s are flavor SU(2) generators.  We used Cho's \cite{cho} Feynman
rules restricted to SU(2), so $\{i,j,k\}\in\{1,2\}$, and $l\in\{1,2,3\}$; the
group theory factor is just the Clebsch-Gordan coefficient $\langle 1,\alpha;
1,\alpha'|0,0\rangle$.  These rules automatically obey unitarity constraints
for $\Lambda\pi\to\Lambda\pi$ analogous to those derived elsewhere \cite{unit}
for $N\pi\to N\pi$.

\INSERTFIG{4.5}{bign_2.eps}{Fig.~2: Single pion emission $\Lambda_b(v) \to
  \Sigma_c(v') \pi(q) W$.}

Since $(\gst/f) \sim \sqrt{N_c}$, the last factor of \puteq{calM} must vanish
at least as fast as $1/\sqrt{N_c}$ when contracted with any final state
$\Sigma^\mu$, which is in turn constrained only by $v'_\mu \Sigma^\mu = 0$:
\Eq \Sigma^\mu (F q_\mu + G v_\mu) = \ord{1\over\sqrt{N_c}} \Endl{1pi}
for any $q$ satisfying $q^2 = m_\pi^2 \approx 0$ (in the chiral limit) and
kinematic constraints.

\subsection{Double Pion Emission}

Double pion emission, $\Lambda_b(v) \to \Lambda_c(v') \pi_l(p) \pi_m(q) W$,
arises from the 3 diagrams of Fig.~3, plus 3 ``crossed'' diagrams related by
$\{l,p\} \leftrightarrow\{m,q\}$.  As long as we restrict our indices to SU(2)
as before, the group theory factor of the crossed diagrams equals that of the
uncrossed diagrams.  Since $(\gst/f)^2 \sim N_c$, the remaining term must
vanish at least as fast as $1/N_c$:
\Eqa 2\eta
  &+& (\eta_1 - w \eta_2) \left[ 2w - {v'\cdot p \over v\cdot p} - {v\cdot p
    \over v'\cdot p} - {v'\cdot q \over v\cdot q} - {v\cdot q \over v'\cdot q}
    \right]
  - \eta_2 \left[ {(v'\cdot p)(v\cdot q) \over (v'\cdot q)(v\cdot p)} +
    {(v'\cdot q)(v\cdot p) \over (v'\cdot p)(v\cdot q)} \right] \nonumber\\
  &+& \eta_1 \left[ {p\cdot q \over (v'\cdot q) (v\cdot p)} + {p\cdot q \over
   (v'\cdot p) (v\cdot q)} \right]
  - \eta \left[ {p\cdot q \over (v\cdot p) (v\cdot q)} + {p\cdot q \over
    (v'\cdot p) (v'\cdot q)} \right]
  = \ord{1\over N_c} \Endla{2pi}

Again, we cannot continue with $n$-pion emission because higher states in the
$I\!=\!J$ tower come into play for $n>2$.

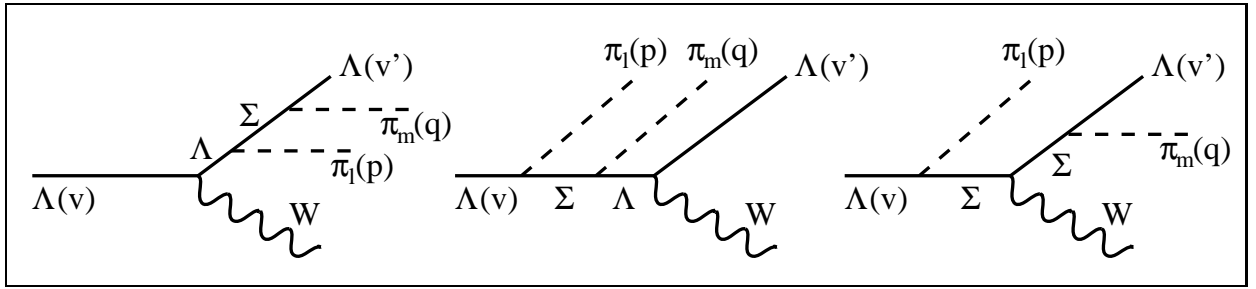
\INSERTFIG{6.49}{bign_3.eps}{Fig.~3: Two-pion emission, $\Lambda_b(v) \to
  \Lambda_c(v') \pi_l(p) \pi_m(q) W$ (3 crossed diagrams not shown).}

\section{Analysis}

\subsection{Taylor Expansion}

Let $\eps^2 = w-1$.  {\it Assume\/} the Isgur-Wise functions can be expanded in
$\eps$; then to $\ord{\eps^2}$,
\Eq \eta(w) = 1 + \eps \et01 + \eps^2 \et02, \qquad
    \eta_1(w) = 1 + \eps \et11 + \eps^2 \et12, \qquad
    \eta_2(w) = \et20 + \eps \et21 + \eps^2 \et22 \End
\Puteq{1lp} becomes
\Eq \eps \left[ 3\et01 - 3\et11 \right] + \eps^2 \left[ 3\et02
  - 3\et12 + 2\et20 - \frac13 \right] + \ord{\eps^3} =
  \ord{1\over N_c} \End
Over different ranges for $\eps$, different terms are constrained:
\Eq \renewcommand{\arraycolsep}{20pt} \begin{array}{rr}
  \hbox{[With $\eps = \ord{N_c^{-3/4}}$]} & \et01 - \et11 =
    \ord{N_c^{-1/4}} \\
  \hbox{[With $\eps = \ord{N_c^{-1/2}}$]} & \et01 - \et11 =
    \ord{N_c^{-1/2}} \\
  \hbox{[With $\eps = \ord{N_c^{-1/4}}$]} & 3\et02 - 3\et12 +
    2\et20 - \frac13 = \ord{N_c^{-1/4}} \end{array} \Endl{tay1}
The latter relation is {\it inconsistent\/} with Chow's result.

Turning to single-pion emission, we go to the $\Sigma_c$ rest frame (where
$\Sigma^0=0$):
\Eq v'=(1,0,0,0), \quad v=(1+\eps^2,0,0,\sqrt2\eps), \quad q\sim (1,
  \sin\theta,0,\cos\theta) \End
and we use the result $\et01 - \et11 = \ord{N_c^{-1/2}}$ from \puteq{tay1}.
Then \puteq{1pi} becomes
\Eq \eps \, (\sqrt2 \sin\theta) \; \vec\Sigma \cdot (\cos\theta,0,-\sin\theta)
  + \ord{\eps^2} = \ord{1\over\sqrt{N_c}} \End
With $\eps = \ord{N_c^{-1/4}}$, this gives a constraint on $\Sigma^\mu$,
representing angular momentum conservation among the light degrees of freedom.
We obtain no information about the $\eta$'s.

We analyze 2-pion emission in the $\Lambda_c$ rest frame, with $p \sim
(1, \sin\bar\theta\cos\bar\phi, \sin\bar\theta\sin\bar\phi, \cos\bar\theta)$
(the normalization of $p$ and $q$ drop out).  Then \puteq{2pi} becomes
\Eqa && \quad \eps \left[ 2(1-p\cdot q)(\et01-\et11) \right] \nonumber\\
     &+& 2\eps^2 \;\Bigl[ \bigl\{ 1 - \cos^2\theta - \cos^2 \bar\theta +
       (\et02-\et12) + 2 \cos\theta \cos\bar\theta \,\et20 \bigr\} \nonumber\\
     && \quad + \; (p\cdot q) \bigl\{ -\cos\theta \cos\bar\theta -
        \frac{1}{\sqrt2} (\cos\theta + \cos\bar\theta) (\et01-\et11) -
        (\et02-\et12) \bigr\} \Bigr] \nonumber\\
     &+& \ord{\eps^3} = \ord{1\over N_c} \Enda
Again using $\et01 - \et11 = \ord{N_c^{-1/2}}$ from \puteq{tay1}, and taking
$\eps = \ord{N_c^{-1/4}}$, we find
\Eq 1 - \cos^2\theta - \cos^2 \bar\theta + (1-p\cdot q)(\et02-\et12) +
  \cos\theta \cos\bar\theta \, (2\et20 - p\cdot q) = \ord{N_c^{-1/4}}
  \Endl{nons}

\subsection{The Exponential Dropoff}

\Puteq{nons}, derived for $\eps = \ord{N_c^{-1/4}}$, cannot be generally true.
We conclude that our assumption of analyticity must be invalid this far from
threshold.  [Nothing significant changes if we try expanding in some other
power of $(w-1)$.]  The Isgur-Wise functions must vanish, \eg\ exponentially,
for $\eps \ge N_c^{-1/4}$, in which case \puteq{2pi} is trivially satisfied.

Indeed, Jenkins, Manohar and Wise \cite{jmw} showed that $\eta \sim
\exp\{-\lambda N_c^{3/2} (w-1)\}$.  So in fact, the Isgur-Wise functions vanish
exponentially fast for $\eps > N_c^{-3/4}$, which is a stronger statement than
ours.

Unfortunately, the only relation we can then retain is the first line of
\puteq{tay1}.  There is no inconsistency with Chow or with kinematics, but
neither can we verify Chow's prediction for $\eta_2(1)$.

\section{Conclusions}

We have analyzed three weak-decay processes ($\Lambda_b \to \Lambda_c W$ at one
loop, $\Lambda_b \to \Sigma_c \pi W$, and $\Lambda_b \to \Lambda_c \pi\pi W$)
in the chiral/heavy/large-$N_c$ limits.  These are the {\it only\/} processes
that do not involve higher states in the $I\!=\!J$ tower.  In this diagrammatic
approach, unitarity requires certain constraints on the baryonic Isgur-Wise
functions.  At threshold, $\eta(1)=\eta_1(1)$ by heavy quark symmetry.  For
$w-1 \approx N_c^{-3/2}$, we still find $\eta=\eta_1$, in agreement with Chow.
The functions vanish exponentially beyond that, and we can derive no further
information.

These unitarity constraints are consistent with, but not as powerful as, Chow's
Skyrme model [or $SU(4)$] relations \cite{chow,chow2}.  In particular,
unitarity constraints give no prediction for $\eta_2(1)$, whereas the Skyrme
model analysis predicts $\eta_2(1)=-1/2$.

We emphasize that we do {\it not\/} disagree with Chow's results.  Rather, we
have shown that perturbative unitarity is incapable of reproducing them.  Here
is an explicit counterexample to the widely-held belief that all large-$N_c$
predictions can be derived from unitarity constraints.

\newpage
\leftline{\Large\bf Acknowledgments} \bigskip
The authors thank C.K. Chow, Ming Lu, Martin Savage, and Mark Wise for helpful
discussions.  This work was partially supported by the U.S. Dept.\ of Energy
under Contract DE-FG02-91-ER40682.

% ======================= BIBLIOGRAPHY ==========================

\def\ap#1{{Ann.\ Phys.} {\bf #1}}
\def\pl#1{{Phys.\ Lett.} {\bf #1}}
\def\np#1{{Nucl.\ Phys.} {\bf #1}}
\def\pr#1{{Phys.\ Rev.} {\bf #1}}
\def\prl#1{{Phys.\ Rev.\ Lett.} {\bf #1}}
\def\prs#1{{Proc.\ Roy.\ Soc.} {\bf #1}}
\def\zp#1{{Z.\ Phys.} {\bf #1}}
\def\ibid{{ibid.\ }}

\end{document}